\documentclass[aps,pra,reprint,superscriptaddress,amsmath,amssymb]{revtex4-2}

\usepackage{graphicx}
\usepackage{hyperref}

\newcommand{\supplement}{Appendix}

\newcommand{\POLI}{Dipartimento di Fisica - Politecnico di Milano, piazza Leonardo da Vinci 32, 20133 Milano, Italy}
\newcommand{\CNR}{Istituto di Fotonica e Nanotecnologie - Consiglio Nazionale delle Ricerche (IFN-CNR), piazza Leonardo da Vinci 32, 20133 Milano, Italy}

\begin{document}

\title{A micro-opto-mechanical glass interferometer for megahertz modulation\\ of optical signals}

\author{Roberto Memeo}
\affiliation{\CNR}\affiliation{\POLI}

\author{Andrea Crespi}
\email{andrea.crespi@polimi.it}
\affiliation{\POLI}\affiliation{\CNR}

\author{Roberto Osellame}
\affiliation{\CNR}

\begin{abstract} 
Waveguide-based interferometric circuits are widely employed in optical communications, sensing and computing applications. In particular, glass-based devices are appealing due to the transparency and bio-compatibility of this substrate, or where low-loss interfacing with fiber networks is required. However, fast electro-optic phase modulation is hard to achieve in glass materials. Here, we demonstrate an optical phase and intensity modulator in glass, working in the megahertz range. This modulator exploits the elasto-optic effect inside a mechanical microstructure, brought to oscillation at resonance, and is entirely realized by femtosecond laser micromachining. In detail, we demonstrate 23-dB optical intensity modulation at 1.17~MHz, with an internal optical loss of the phase-modulator component as low as 0.04~dB.
\end{abstract}

\maketitle
\section{Introduction}

In the last decades, integrated photonics has played an increasing role in the development of new technologies and devices, with applications in the field of optical communications, sensing, and more recently also in linear-optical quantum computing and simulations. The monolithic integration of a multitude of waveguide components inside a minute optical chip is indeed the key to build ultra-stable interferometric circuits for large-scale processing of classical and quantum optical signals \cite{OBrien2009, Bogaerts2020}.

{Different applicative scenarios involve different strategies to act dynamically on the device function. Slow or quasi-static reconfiguration of the device function, which is of interest for instance in many quantum information applications,} is often achieved by exploiting the thermo-optic effect. Electrical microheaters deposited above the waveguide produce a localized temperature increase, which can dynamically modulate the refractive index and thus the optical phase of the propagating light. This method has the advantage of technological simplicity and of being viable in practically any material platform, however it suffers from important limitations in terms of modulation bandwidth. In fact, the response time of thermo-optic phase shifting is typically in the order of {millisecond or several} microseconds {at most}, which is dictated by the thermal diffusion dynamics \cite{Harris2014, Calvarese2022}.

{On the other hand, rapid switching and modulation functions may be required e.g. for signal encoding, multiplexing and demultiplexing. These functions are typically implemented by exploiting material-specific features,} such as carrier injection or the electro-optic effect. The former requires a semiconductor optical substrate, while the second one typically needs a medium with sufficiently high nonlinear response. As a prominent example, electro-optic devices in lithium niobate find nowadays widespread application in the optical communications field, achieving response times  in the order of 100~fs \cite{Zhang1997} and high modulation frequency capabilities, up to 100~GHz \cite{Weigel2018, WangNat2018, Wang2019, He2019}. However, the refractive index mismatch between crystalline or semiconductor substrates (where the modulator is built) with respect to the silica glass of fiber connections cause almost unavoidable losses at the interfaces, thus increasing the insertion losses of commercial fiber-coupled modulators well above 3 dB. 

Low-loss interfacing with optical fibers is more easily achieved with glass-based waveguide devices. Reconfigurable optical circuits in silica glass can be realized both with lithographic methods \cite{Takahashi2011, Carolan2015} and with direct writing techniques \cite{Flamini2015}. In particular, reconfigurable glass circuits fabricated with the femtosecond laser writing technology have recently demonstrated their advantages in several applications, including quantum information \cite{spagnolo2022,skryabin2023,cimini2023}, microrheology\cite{Vitali2020}, and integrated microscopy\cite{Calvarese2022opt}.
{However, glass does not show, intrinsically, a significant electro-optic response. Thermal poling has been adopted in the literature to introduce a second-order nonlinearity both in lithographic \cite{abe1996} and in femtosecond-laser written \cite{li2006,an2014,ng2016} waveguides in glass. This however adds a nontrivial processing step to the fabrication. A proof of principle experiment was also reported \cite{Humphreys2014}, in which an external mechanical actuator was employed to induce a local strain in a waveguide inscribed by femtosecond laser pulses, and thus produced a phase shift with quick time response through the elasto-optic effect. As a matter of fact, nowadays the application of glass-based waveguide circuits is commonly restricted to the cases where the slowness of thermo-optic phase actuation is tolerated.}

\begin{figure*}[t!]
    \centering
    \includegraphics[width=0.8\textwidth]{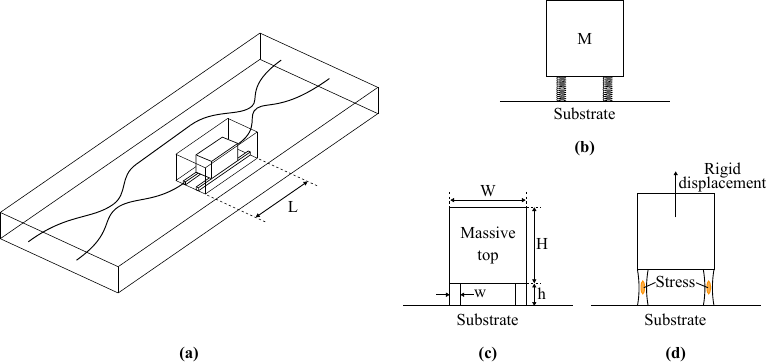}
    \caption{\textbf{(a)} Scheme of the Mach-Zehnder interferometer, with the resonant three-dimensional microstructure enclosing one of the arms. \textbf{(b)} Concept of the mechanical resonator: a massive top suspended on two elastic supports. \textbf{(c)} Cross-section of the microstructure implementing the resonator: the relevant dimensional parameters are indicated. \textbf{(d)} Qualitative illustration of the effect of a displacement of the massive top, causing a mechanical stress into the supporting rails.}
     \label{Cube scheme image}
\end{figure*}

Here we present a novel approach to modulate optical signals in glass waveguides, with frequencies in the order of the megahertz, relying on the elasto-optic effect. In detail, we propose and demonstrate in experiments a phase modulator based on the resonant oscillations of a mechanical microstructure, embedding a laser-written optical waveguide. A periodic mechanical stress is thus applied to the waveguide during the oscillation, which in turn causes a refractive index modulation due to the elasto-optic effect. The resulting phase modulation is transformed into signal switching by enclosing the modulator into a waveguide Mach-Zehnder interferometer (MZI), as shown in Fig.~\ref{Cube scheme image}a. The entire device is realized in fused silica glass by femtosecond laser micromachining (FLM).

\section{Materials and methods}

FLM is based on the permanent material modification induced by irradiation with femtosecond laser pulses, focused by suitable optics \cite{Osellame2012}. The substrate is ordinarily transparent to the laser wavelength, thus absorption of the laser pulse only occurs very close to the focus, where the high peak intensities are able to trigger nonlinear interaction processes. By properly tuning the irradiation parameters, smooth refractive index modifications can be achieved, and optical waveguides are literally drawn in the substrate by controlling its relative translation with respect to the laser focus. On the other hand, tracks irradiated by pulses with higher energies show larger sensitivity to chemical etching, and substrate microstructuring can be accomplished by selective removal of the irradiated regions \cite{Osellame2011, Memeo2021}.

To fabricate our device we employ a commercial Ytterbium-based femtosecond laser system (Light Conversion Pharos), which delivers 170-fs pulses at a wavelength of $\sim$1030~nm, with a maximum average output power of 10~W. Both the waveguides and the tracks to be etched are irradiated in the same session (though with different parameters) and using the same focusing objective (Zeiss N-Achroplan) with 0.45 NA. This guarantees an accurate alignment between the waveguide circuits and the mechanical microstructure. We adopt commercial grade fused silica (JGS1, Foctek Photonics Inc.) as a substrate. High-accuracy translation of the sample below the laser beam is performed by means of computer controlled linear stages (Aerotech ABL1500), which combine air-bearings and brushless servomotors to guarantee smooth displacements and movement precision in the order of 10 nm. 

In detail, the regions to be removed, in order to define the mechanical micro-resonator, are irradiated with a pulse energy $E_P$~=~1.5$\mu$J, pulse repetition rate of 500~kHz, and using a translation speed of 2~mm/s. Laser polarisation is always kept orthogonal to the translation direction to achieve the maximum etching selectivity \cite{Hnatovsky2006, Yu2011}. Optical waveguides are then inscribed with a pulse energy $E_P$~=~ 220~nJ, pulse repetition rate of 100~kHz and translation speed of 1~mm/s. The cross-section of each waveguide is defined by {20 partially overlapped laser scans, each laterally shifted from the previous one by 0.4~$\mu$m} \cite{Nasu2005, Psaila2006}.

Waveguide irradiation parameters are optimized to support a single spatial mode, of almost circular shape, at the operating wavelength of 1550~nm. In detail, the mode diameter at this wavelength is about 11.5~$\mu$m ($1/e^2$), which results in a typical coupling loss with optical fibers (at the terminal facets of the chip) of 0.1~dB/facet; propagation loss is experimentally estimated to be 0.44~dB/cm. The waveguides exhibit intrinsic modal birefringence of about $1.2 \cdot 10^{-5}$, with a fast axis oriented vertically, i.e. along the propagation direction of the writing laser beam.

After laser irradiation, we perform a two-step chemical etching process \cite{LoTurco2013}. First, we immerse the glass sample for 20 minutes in aqueous solution of hydrofluoric acid (20\% volume concentration) kept at 35$^\circ$C temperature. The high etching rates achieved with hydrofluoric acid allows rapid material removal, at the price of a low precision in the details. In fact, we then perform a second etching step, immersing the glass sample for 24 hours in aqueous solution of potassium hydroxide (10 M concentration), kept at 90$^\circ$C temperature. Thanks to the higher selectivity of potassium hydroxide, together with its slower action, this second step enables a more precise etching of minute details of the micro-resonator. This is crucial especially in the regions that contain the inscribed optical waveguides, as an imprecise material removal could result in waveguide damage.

The micro-mechanical oscillators demonstrated in this work are driven by vibrations of the whole substrate, excited by a piezoelectric actuator  (000023074, PI Ceramic).
In detail, we use a metallic clamp to fix the glass sample between the piezo-actuator, which is a millimeter-size ceramic disk {with 5-mm diameter and 2-mm thickness}, and a spring.
The piezo-component is electrically actuated by a function generator (Tektronix AFG3011C), providing a variable voltage signal with a maximum output peak-to-peak amplitude of 40~V. When a greater voltage signal is needed, a custom-made 3$\times$ passive voltage transformer is applied at the output of the function generator. 

\section{Device design and experimental results}

\subsection{Design and optimization of the micro-mechanical resonator} \label{sec:desOpt}

To design our optically-coupled mechanical resonator, we take inspiration from the archetypical mass-spring oscillator. There, the mechanical strains and stresses are concentrated in one part of the system (the 'spring') while the other part (the 'mass') shows little intrinsic deformation during the motion, and governs the inertia of the system. In our scenario, we need to carve both the 'spring' and the 'mass' as different parts of a homogeneous glass structure: the former should be thin and deformable, the latter more massive and rigid. The waveguide would be inscribed into the more strained region, where the stress-induced birefringence would produce a local change in the propagation constant of the guided mode, which eventually results in a phase shift of the propagating light. Importantly, in order to obtain clean phase modulations without spurious polarization rotations, the principal direction of the mechanical stress (which coincides with the axis of the induced birefringence) should be as parallel as possible to the intrinsic vertical axis of the waveguide birefringence.

From these considerations, we come to the design illustrated in Fig.~\ref{Cube scheme image}.  It consists of a massive top with an approximately square cross-section, connected to the substrate by two thin rails. A vertical displacement of the top part would produce a deformation and a consequent stress mainly localized on the two elements beneath, as shown in Fig.~\ref{Cube scheme image}d. In other words, the top part acts as the 'mass' of the oscillator while the two rails act as two symmetric 'springs'.
An optical waveguide would be inscribed in one of these stressed regions, in order to obtain the desired elasto-optic phase modulation if the 'up-and-down' oscillation mode is excited. Note that the symmetric design, with two springs instead of one, should make the device more robust and less prone to oscillate laterally as an inverted pendulum.

A trivial physical modelling, which considers the top part as rigid and the elastic deformation localized in the rails, predicts a resonance frequency  $f$ for the oscillation mode of interest that reads:
\begin{equation}
f = \frac{1}{2\pi} \sqrt{\frac{k_{eq}}{M}} =  \frac{1}{2\pi} \sqrt{\frac{2 E}{\rho WH} \, \frac{w}{h}} \label{eq:resFreq}
\end{equation}
where  $k_{eq}$ is the equivalent spring constant and $M$ is the mass of the top part. The expression has been developed taking into account the elastic modulus $E$ and the density $\rho$ of the material, as well as the geometrical dimensions of the microstructure (as indicated in Fig.~\ref{Cube scheme image}c), namely substituting:
\begin{align}
k_{eq} &= 2E \frac{w L}{h}  & M &= \rho W H L
\end{align}
where we have further indicated as $L$ the length of the device parallel to the waveguide propagation direction (dimension orthogonal to the cross-section represented in the figure). Note that the resulting frequency $f$ does not depend on $L$. 

{However, the length $L$ must be chosen judiciously, so that the required phase shifts are produced with an amount of stress compatible with the mechanical limits of the substrate. To produce a full switch of an optical signal in a MZI, a phase modulation of $\pm \pi/2$ is required, which corresponds to a refractive index modulation $\pm \Delta n = \pm \frac{\lambda}{4 L}$. The refractive index change is proportional to the applied stress $\sigma$ according to $|\Delta n| =  c | \sigma |$; thus, we write the required stress: 
\begin{equation}
\sigma_{\pi/2} = \frac{\lambda}{4 c} \frac{1}{L}
\end{equation}}

{In addition, for a given force $F$ applied to the massive top in the vertical direction, the rails are vertically deformed by $\Delta h = F / k_{eq}$. The strain of the basis rail is:
\begin{equation}
\frac{\Delta h}{h} = \frac{F}{k_{eq}} \frac{1}{h} = \frac{F}{2 E L} \, \frac{1}{w}
\end{equation}
Thus, in order to efficiently produce stress on the waveguides (which is proportional to the strain), one needs thin basis rails. On the other hand, the dimension $w$ needs to be chosen as sufficiently larger than the waveguide mode size (to avoid additional optical losses) and much larger than the tolerance of the fabrication process (hundreds of nanometers at best). Otherwise, one would risk to observe poor agreement between the predicted natural frequency and the experimentally measured one. Additional constraints on the dimensions are given by the performance of the laser-assisted etching technique in terms of aspect ratios of the hollow parts \cite{LoTurco2013}. 

As a compromise between these different needs, we can consider for instance} 
$w$~=~50~$\mu$m, $h$~=~100~$\mu$m{, $W = H =$~500~$\mu$m, and $L$~=~1 mm, for which} Eq.~\eqref{eq:resFreq} gives a resonance frequency of about 1.8~MHz. {Considering a stress-optic coefficient for fused silica $c = 4.5 \cdot 10^{-12}\; \mathrm{Pa}^{-1}$ \cite{Bellouard2011}, for light polarized orthogonally to the main stress direction, we get  $\sigma_{\pi/2} = 86\;\mathrm{MPa}$. Note that microstructured fused silica can withstand very high compressive and tensile limits without damage, even above 1~GPa \cite{Bellouard2011}.
These results shows that modulators with resonance frequencies in the megahertz range should be at reach} with a structure having {dimensions} within the possibilities of our micromachining technology.

\begin{figure*}
    \centering
    \includegraphics[width=0.65\textwidth]{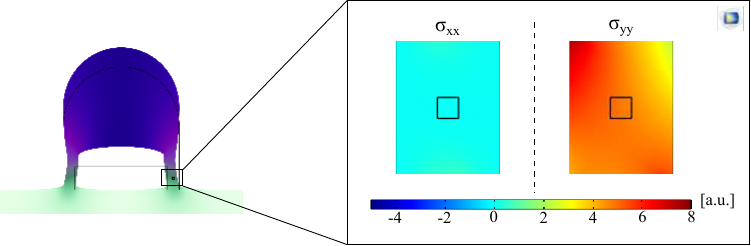}
    \caption{Stress states across the rectangular basis 'rails' of the rounded micro-mechanical resonator. The solid black square in the inset shows the position of the optical waveguide.}
    \label{Cube stress image}
\end{figure*}

Fine tuning of the oscillator design was accomplished by harnessing numerical simulations of the mechanical vibration eigenmodes, performed with COMSOL Multiphysics.
Fig.~\ref{Cube stress image} shows {an example of a }simulated eigenmode shape, for the 'up-and-down' resonance, together with the associated stress states across the resonator basis 'rails'.
In particular, such simulations showed that it is not possible to neglect the deformation of the massive top and of the substrate during the oscillations, which decrease the effective elastic constant and lowers the natural oscillation frequency with respect to the simple model of Eq.~\ref{eq:resFreq}. For instance, the same aforementioned geometry, for which Eq.~\eqref{eq:resFreq} predicts an oscillation frequency of 1.8~MHz, yields an oscillation frequency of 1.18~MHz when simulated numerically.

\begin{figure*}
    \centering
         \includegraphics[width=0.8\textwidth]{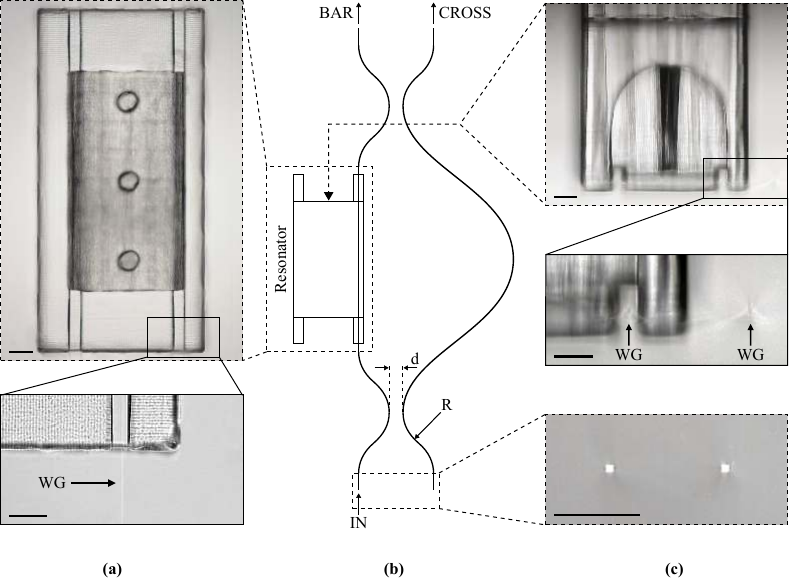}
    \caption{\textbf{(a)} Focus-stacking picture of the micro-resonator as seen from above: the inset show the position of the optical waveguide within the supporting rail. \textbf{(b)} Geometrical scheme of the unbalanced MZI with the quantities of interest. \textbf{(c)} Microscope picture of the resonating structure as seen from the side: insets show the optical waveguides of the two branches of the MZI within the bulk substrate and at the input facet. Scalebars correspond to 100~$\mu$m.}
      \label{Cubic resonator image}
\end{figure*}

{In a set of preliminary experiments, several microresonators with the above-described geometrical shape were fabricated using the procedure discussed in the Methods; we experimented slightly different dimensional parameters and geometrical details, while keeping the nominal resonance frequency to about 1-1.1~MHz. In these devices s}traight optical waveguides were inscribed in the centers of the basis 'rails', as seen in the inset of Fig.~\ref{Cubic resonator image}. {T}hese experiments allowed to verify that the waveguide passing in the microstructure do not experience measurable additional optical loss with respect to a standard waveguide inscribed in the bulk glass. {Secondly}, they allowed to test the behaviour of the mechanical oscillator. 

To the latter purpose, the glass sample was fixed against the piezo-actuator, as described in the Methods, and diagonally polarized light was coupled in the waveguide while driving the piezo with a sinusoidal voltage. By analysing the dynamical polarization rotations, induced on the light propagating in the waveguide for different frequencies and amplitudes of the driving signal, we were able to assess the operation principles of our devices. We could verify a good correspondence between the experimental resonance frequencies and the ones predicted by the numerical simulations. In addition, we could ascertain that during the oscillation the birefringence axis of the waveguide remain{ed} vertically oriented without rotating significantly, which indeed enables to employ our micro-opto-mechanical device inside a Mach-Zehnder interferometer (see the \supplement for further details on these experimental measurements).

{Most efficient optical modulation was accomplished for microresonators with the design reported in Fig.~\ref{Cube stress image}; nominal dimensions are $w$~=~50~$\mu$m, $h$~=~100~$\mu$m, $W = H =$~500~$\mu$m, and $L$~=~1~mm. In addition, the top of the resonating structure was rounded and three cylindrical holes, with a diameter of 100~$\mu$m, have been etched traversing the resonator from top to bottom, as shown in Fig.~\ref{Cubic resonator image}a. The holes reduce possible action of air friction and facilitate the access of the etchant solution inside the cavity below the massive top, during the selective-etching process, thus allowing for a more accurate definition of the resonator shape.}

\subsection{Integrated elasto-optic phase modulator}

In order to obtain a fully-integrated device exploiting the resonant modulation, we inscribed the micro-resonating structure on one arm of an integrated MZI. The induced index change is responsible for a phase shift between the signals propagating inside the two branches of the MZI, resulting in a modulation of the transmitted light at the two output ports. Moreover, the MZI was designed with an unbalanced layout, yielding an optical path difference of 11~$\mu$m between the two arms. In this way, the working point can be tuned by changing the input wavelength. 

Details of the interferometer layout are reported in Fig.~\ref{Cubic resonator image}b. Bent waveguide segments are circularly curved with a radius (R) of 50~mm, and the waveguides in the directional couplers are brought close at 11.0~$\mu$m distance (d), to achieve balanced splitting at the central wavelength of 1550~nm (for horizontal polarization).

To characterize the device operation, we clamped the substrate against the piezo-actuator as described in the Methods, and drove the actuator with a sinusoidal voltage signal. Horizontally polarized light from a tunable laser source was coupled in one of the input waveguides, while the transmitted light from both outputs was collected by a microscope objective and focused onto two distinct amplified photodectors, based on InGaAs photodiodes (Thorlabs PDA20CS2). The photodetector signals are monitored by an oscilloscope. 

In this setting, we can study the device transmission at the bar or cross output states as a response to a sinusoidal voltage signal fed to the piezo $\Delta V(t) = V_0 \sin (2 \pi f t)$. A simple model for a Mach-Zehnder interferometer with balanced losses in the two arms, and built of identical beam splitters, gives:
\begin{equation}
    T_\mathrm{C} = T_0 \, \cos^2 \Bigl[\alpha(f) \, V_0 \sin \bigl(2 \pi f t + \theta(f)\bigr) + \phi_0\Bigr]
    \label{eq:T}
\end{equation}
being $T_\mathrm{C}$ and $T_\mathrm{B} = 1 - T_\mathrm{C}$ the normalized transmissions for the device cross and bar states respectively. The modulation coefficient $T_0$ depends on the input wavelength $\lambda$, on the splitting ratio of the couplers, and on the optical losses of the device.
$\alpha(f)$ and $\theta(f)$ represent the amplitude and phase of a transfer function, which transforms the oscillation of the piezo into the oscillation of the optical phase; they depend on the characteristic features of the micro-structure and of the actuation process, as well as on the frequency $f$. Finally, $\phi_0$ is an intrinsic phase term of the MZI ring. In our setting, by changing the input wavelength we act mainly on the phase $\phi_0$ and, less relevantly, on the splitting ratio of the directional couplers composing the MZI.

\begin{figure}
    \centering
    \includegraphics[width=\linewidth]{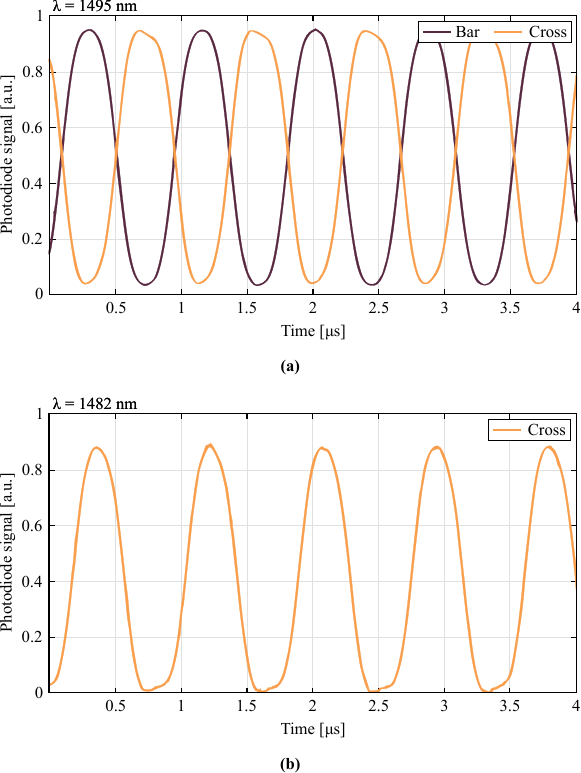}
    \caption{\textbf{(a)} Modulated signals at the two output ports of the integrated MZI.
    \textbf{(b)} Output signal at the cross state when the device operates as a modulator. The driving voltage is sinusoidally varied with a peak-to-peak amplitude of 79~V at $f$~=~1.17~MHz. Signals are normalised over the maximum transmitted power.}
    \label{MZI measure image}
\end{figure}

First, by scanning the driving frequency we can experimentally identify the resonance frequency of this device, which is the one for which $\alpha(f)$ is maximum, found to be 1.17 MHz (within a 1\% range from the simulated value). The input wavelength was then tuned to achieve a modulation around  $\phi_0 = 0$. Our source has a tuning range of about 100~nm around a central wavelength of 1530~nm; the optimal working point was experimentally found at 1495~nm wavelength.
Finally, we increase the driving voltage up to the achievement of the maximum contrast between peaks and dips of the interference fringes; namely we maximize the extinction ratio (ER):
\begin{equation}
\text{ER}|_{dB} = 10 \, \log \left(\frac{T_{max}}{T_{min}}\right)
\label{eq:er}
\end{equation}
Fig.~\ref{MZI measure image}a shows the measured signals, normalized on the maximum transmitted power ($T_\mathrm{C} + T_\mathrm{B}$). In detail, we measure the following ERs for the bar and cross signals: $\text{ER}_\text{bar}$~=~15.8~dB, $\text{ER}_\text{cross}$~=~14.1~dB.
These values for the ERs are limited by the non-ideal splitting ratio of the MZI couplers, which worsens if the input wavelength deviates significantly from the design one of 1550~nm.

In this configuration, the device is operating as a switch, periodically directing the input power towards one of its two different outputs. However, by tuning the input wavelength one can set the working point of the device such that the cross state results completely depleted in a periodic fashion. In fact, it is always possible to find an operating regime for which the cross transmission periodically tends to zero (see Eq.~\eqref{eq:T}). In our experiments, we found this regime at about 1482~nm wavelength: the resulting cross signal is reported in Fig.~\ref{MZI measure image}b {(a graph showing also the bar signal is presented in Fig.~S4 of the \supplement)}. In this condition, the device behaves like an intensity modulator with a measured extinction ratio $\text{ER}_\text{mod}$~=~23.2~dB.

Finally, we can estimate the quality factor (Q) of the micro-mechanical resonator by analysing the response of the device around resonance; for the considered device, we obtain Q~$\sim$~450. {See the {\supplement} for further details.}

\section{Conclusions}

We have demonstrated an integrated optical intensity modulator and switch, fabricated by FLM in glass as a monolithic device. A mechanical microstructure and an optical interferometer are realized with perfect relative alignment in the same substrate: resonant oscillations of the microstructure provoke a periodic mechanical stress on the waveguide, which induce a phase modulation by the elasto-optic effect. Note that in this device the phase modulation occurs in a 1-mm long waveguide segment, with an associated internal loss of about 0.04~dB.

The resonance frequency of the mechanical oscillator can be tuned by changing its geometrical dimensions, going well beyond the demonstrated modulation at 1.17~MHz thanks to the spatial resolution of {FLM. In addition, one could think in the future to engineer resonators able to exploit several different oscillation eigenmodes, with different frequencies, and able to work in different operating conditions. 

T}he limits of the bulk actuation setup and the piezo-electric disk specifications did not allow us to reach frequencies higher than those here reported.
{T}he necessity of using driving voltages in the order of 80~V, despite the relatively high quality factor, {also} suggests that in our setting the excitation of the sample vibrations by the piezo is not efficient.
For these reasons, actuation by micro-piezos, deposited on the top-surface of the chip, is envisaged as a future development. 

This solution would allow not only an efficient actuation of the resonant modulation, but also quasi-static phase tuning. {For our device, a vertical static force of about 17~N exerted on the resonator top would be sufficient to produce a $\pi$ phase shift, as can be readily evaluated from the physical model discussed in Section~\ref{sec:desOpt}. However, an optimization of the device shape for this operating condition could further improve the static actuation efficiency.} Elasto-optic modulation with localized actuators {would indeed conjugate} a rapid time response with negligible cross-talks to adjacent devices. 

While a similar technology has been recently demonstrated in silicon nitride \cite{dong2022}, the realization of elasto-optic modulators in glass may open the door to a new class of devices, which could combine efficient coupling with optical fibers and functional interplays with microfluidic networks \cite{Vitali2020,Calvarese2022opt}.

\section*{Acknowledgments}
This work is supported by the European Union's Horizon 2020 research and innovation programme under the PHOQUSING project GA no. 899544. AC acknowledges funding by the PRIN2017 programme of the Italian Ministry for University and Research, QUSHIP project (Id. 2017SRNBRK).

\section*{Disclosures}
The authors declare no conflicts of interest.

\section*{Data availability}

Raw data of the graphs reported in the figures are publicly available on Zenodo (\url{https://doi.org/10.5281/zenodo.11354220}). Further data underlying the results presented in this paper may be obtained from the authors upon reasonable request.


\bibliography{Bibliography}

\clearpage
\appendix
\renewcommand\thefigure{A\arabic{figure}}
\setcounter{figure}{0}
\renewcommand\theequation{A-\arabic{equation}}
\setcounter{equation}{0}

\section*{Appendix}

\subsection{Investigation on the induced birefringence modulation}

In order to assess the device capabilities of producing a localised refractive index variation through the elasto-optic effect, we fabricated several micromechanical resonators containing one straight waveguide, as discussed in the Main Text. In detail, the characterization procedures involves measuring the polarization rotation induced by the oscillation of the resonator on the light propagating in the waveguide.

\begin{figure*}
    \centering
    \includegraphics[width=0.7\textwidth]{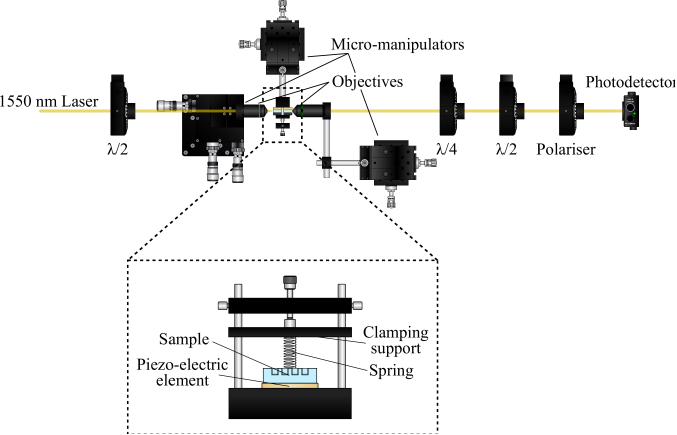}
    \caption{Schematic representation of the characterisation setup adopted to assess the induced birefringence modulation due to the micro-resonator vibrations.}
    \label{Cube characterisation setup image}
\end{figure*}

The glass chip containing the device is clamped onto a custom-assembled support (see the Methods section), and mounted in the characterisation setup depicted in Fig.~\ref{Cube characterisation setup image}. Linearly polarized coherent light from a diode laser at 1550~nm wavelength is injected in the  waveguide by means of a 10$\times$/0.25~NA microscope objective, and then collected at the output and focused onto a fast InGaAs photodetector by a 25$\times$/0.50~NA objective. Both the chip and the focusing optics are mounted on 3-axis micromanipulators.

A half-wave plate is used to rotate the laser polarisation at the device input, whereas the quarter-wave and half-wave plates before the detector, together with a linear polariser, are adopted to project the output state onto two orthogonal polarization bases.

The described setup is actually equivalent to a polarisation interferometer. In particular, when a superposition of H and V polarisation states (the + state) is coupled to the waveguide, the vibration of the resonator induces a periodic birefringence variation, resulting in a dephasing of the two polarization components (provided that the direction of the birefringence axis is not changed). 
The linear polariser then recombines the polarisations at the output (at +45$^\circ$ or -45$^\circ$) revealing a signal modulation that depends on the induced phase shift. This is indeed equivalent to an MZI where only one of the two input polarisation components is subject to a phase shift. The other output waveplates are adopted to add an arbitrary phase term between H and V, allowing to set the working point of the device in the middle of the interference fringes, so that the output signals are symmetric. 

The oscillation of the microstructure is obtained by forcing an oscillation of the whole glass chip, using a piezoelectric disc actuator. The frequency of the sinusoidal driving voltage is swept around the simulated value for the device resonance frequency, while keeping the amplitude constant to 40~V peak-to-peak. The experimental natural oscillation frequency of the micro-resonator is identified as the frequency which provides a non-negligible modulation of the output optical signal.  

Once the optimal excitation frequency is found, a voltage transformer can be inserted between the function generator and the piezo, to increase the amplitude of the driving signal up to 112~V peak-to-peak.

\begin{figure*}
    \centering
    \includegraphics[width=0.6\linewidth]{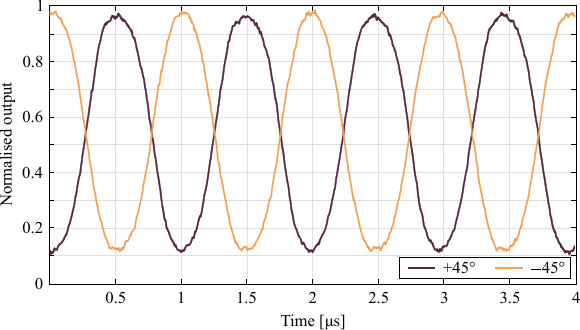}
    \caption{Signal modulation at the output of the rounded "brick-shaped" micro-resonator over two orthogonal polarisation bases. The driving voltage is sinusoidally varied with a peak-to-peak amplitude of 112~V at $f = 1.024\;\text{MHz}$. Signals have been normalised to their maximum value.}
    \label{Cube modulation image}
\end{figure*}

We report in Fig.~\ref{Cube modulation image} an example of the acquired signal, for a device resonating at 1.02~MHz (the angles of the output waveplates are optimised to achieve the same average intensity for both signals).

\begin{figure*}
    \centering
    \includegraphics[width=0.9\linewidth]{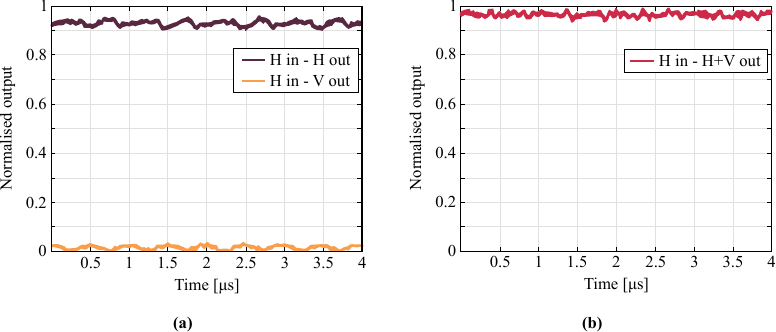}
    \caption{\textbf{(a)} Residual modulation when an H polarisation state is injected in the device and projected at the output on the H (red) and V (yellow) states. \textbf{(b)} Cumulative (H+V) power transmitted by the device. All signals have been normalised to their maximum value.}
    \label{Verification measure image}
\end{figure*}
Signals are symmetric and show a good visibility of above 10~dB (see Eq.~4 in the Main Text).

It is worth noting that the resonant actuation of the device is performed by exciting the vibration of the whole glass sample. This, in principle, may induce a decoupling of the device waveguides due to a relative motion of the substrate with respect to the focusing optics, which is an undesirable effect since it perturbs the phase modulation measurement. In order to exclude that 
the observed optical modulation is due to periodic decoupling of the waveguide, additional measurements are needed.
As well, possible spurious rotations of the birefringence axis under the generated stress states should be investigated.

To these purposes, H polarised light is injected into the device and the output state is analysed by rotating the polariser in the H and V configurations. If a rotation of the birefringence axis occurs during the device resonant operation, a fraction of the transmitted H input signal should be periodically transferred to the V state at the output. A periodic decoupling would be instead understood from modulations of the sum of the H and V signals (which amounts to the total transmitted power).

As reported in Fig.~\ref{Verification measure image}(a), a slight modulation is visible - actually negligible with respect to the one obtained in the experiment of Fig.~\ref{Cube modulation image}. This modulation is not visible on the sum of the H and V signals (Fig.~\ref{Verification measure image}(b)).

These results confirms that periodic decoupling is not playing a role in the signal modulation reported in Fig.~\ref{Cube modulation image}, and that the modulation of the birefringence during the oscillation can be described with good accuracy as a modulation on the birefringence value, while the axis is kept mostly fixed and aligned vertically. Namely, the results in Fig.~\ref{Cube modulation image} are genuinely related to a phase shift between the polarisation components. 

As a final note, straight optical waveguides have been also inscribed in the bulk substrate, far away from the resonator. When exciting the sample vibrations they do not show any modulation of the transmitted signal, strengthening the validity of the obtained results.


\subsection{Quality factor of the resonating micro-structure}

In the Main Text, we show that the proposed resonating micro-structure is designed in analogy with a simple mass-spring system. As a result, its dynamic behaviour and its response at resonance can be approximately modelled by an equivalent one-dimensional harmonic oscillator. In our particular case, the motion of the mass is triggered by an external piezo-actuator which excites a vibration of the whole sample; it thus corresponds to a mass-spring system actuated by its base and not by applying a force to the mass. In this configuration, when the system is kept within the linear regime and the amplitude of the driving oscillations is not too large, the periodic motion of the sample substrate, driven by the deformation of the piezo-electric element, is transferred to the displacement of the mass through a complex transfer function of the kind \cite{spagnolo2020}:
\begin{equation}
    T(f) = T_p \frac{f \left( f - j \dfrac{f_0}{Q} \right)}{\left( f_0^2 - f^2 \right) + j \dfrac{f_0}{Q} f}
    \label{eq:T(f)}
\end{equation}
being $f_0$ the resonance frequency and $Q$ the quality factor. $T_p$ describes the coupling between the piezo deformation and the substrate oscillation, which is assumed constant with $f$ nearby $f_0$. 

The oscillation of the micro-structure is directly linked to a deformation of its supporting rails, where one branch of the unbalanced MZI is inscribed. In addition, there is a linear dependence between the deformation of the rails and the induced phase shift due to the elasto-optic effect. If the actuation voltage of the piezo is kept low enough and the wavelength is correctly set, the relation between the optical modulation at the output of the MZI and the phase modulation is approximately linear.

Thus, to experimentally retrieve $Q$, we measured the amplitude of the modulation of the optical signal while sweeping the driving frequency of the piezo, and fitted the experimental data with the modulus of  Eq.~\eqref{eq:T(f)}, multiplied by a proper constant. 

\begin{figure*}
    \centering
    \includegraphics[width=0.9\textwidth]{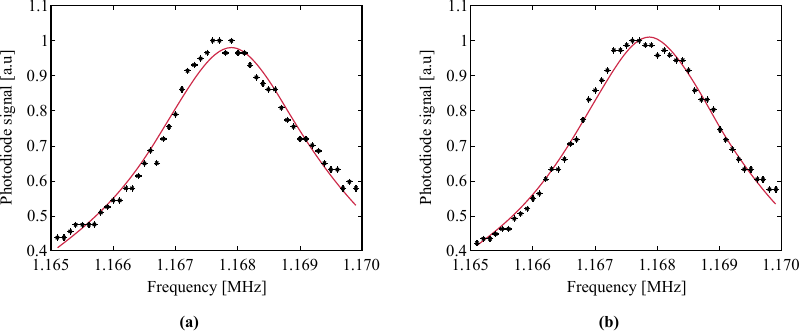}
    \caption{Frequency response of the mechanical oscillator \textbf{(a)} at ambient pressure and \textbf{(b)} in vacuum ($p = 4 \cdot 10^{-3}$~mbar). {To give a quantitative reference, one can remind that at the resonance peak a $\pm \pi/2$ phase modulation (see Fig.~4 in the Main Text) is achieved with a sinusoidal excitation voltage of about $\pm$40~V on the piezo. This corresponds to a peak transfer coefficient between excitation voltage and phase modulation of 0.039~rad/V.}}
    \label{Q image}
\end{figure*}

In detail, we first tuned the input wavelength to achieve a modulation around $\phi_0 = 0$, as in the case of the switch operation (see the Main Text); we then finely swept the frequency of the sinusoidal driving voltage of the piezo-actuator around the resonance, while keeping the actuation voltage fixed at 12~V peak-to-peak, and we recorded the amplitude of the output signal on the cross state. The obtained experimental curve gave a best fit with $Q \simeq 450$ at ambient pressure. 
We repeated the same measurement in a vacuum environment ($p = 4 \cdot 10^{-3}$~mbar), obtaining $Q \simeq 460$. Fig.~\ref{Q image} shows the measured signal amplitude at the cross state both in air and in vacuum, with varying $f$, together with its best fit with a function of the kind of Eq.~\eqref{eq:T(f)}. 

Note that these experimental data indicate that the quality factor of the mechanical oscillator does not change significantly when operating in vacuum, suggesting that air friction is not the main limiting factor. Instead, the quality factor is possibly limited in this device by other dissipation channels, such as thermo-elastic losses due to material viscosity.

\begin{figure*}
    \centering
    \includegraphics[width=0.6\linewidth]{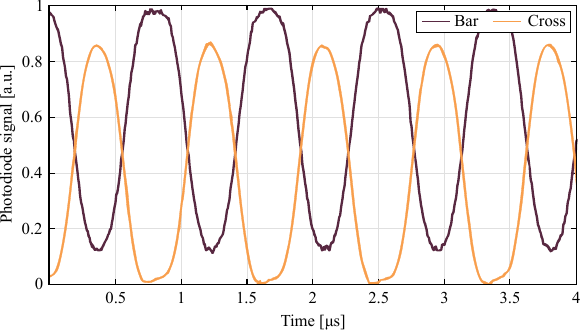}
    \caption{{Output signals (bar and cross) when the device operates as a modulator, namely in the experiment shown in Fig.~4b of the Main Text. The driving voltage is sinusoidally varied with a peak-to-peak amplitude of 79~V at $f$~=~1.17~MHz.}}
    \label{barCross}
\end{figure*}

\end{document}